\documentclass[numberedappendix]{emulateapj}
\usepackage{apjfonts}
\usepackage{hyperref}
\newcommand{\rs}{r_{\rm S}}

\shorttitle{The Properties and Fate of the G2 Cloud}
\shortauthors{Shcherbakov}

\begin{document}
\title{THE PROPERTIES AND FATE OF THE GALACTIC CENTER G2 CLOUD}
\author{Roman V. Shcherbakov\altaffilmark{1,2,3}}
\email{roman@astro.umd.edu}
\altaffiltext{1}{\url{http://astroman.org} \linebreak Department of Astronomy, University of Maryland, College Park, MD 20742, USA}
\altaffiltext{2}{Joint Space Science Institute, University of Maryland, College Park MD 20742, USA}
\altaffiltext{3}{Hubble Fellow}

\begin{abstract}
The object G2 was recently discovered descending into the gravitational potential of the supermassive black hole (BH) Sgr A*.
We test the photoionized cloud scenario, determine the cloud properties, and estimate the emission during the pericenter passage.
The incident radiation is computed starting from the individual stars at the locations of G2.
The radiative transfer calculations are conducted with CLOUDY code and $2011$ broadband and line luminosities are fitted.
The spherically symmetric, tidally distorted, and magnetically arrested cloud shapes are tested with both the interstellar medium dust and $10$~nm graphite dust.
The best-fitting magnetically arrested model has the initial density $n_{\rm init}=1.8\times10^5{\rm cm}^{-3}$, initial radius $R_{\rm init}=2.2\times10^{15}{\rm cm}=17 {\rm mas}$,
mass $m_{\rm cloud}=4M_{\rm Earth}$, and dust relative abundance $A=0.072$.
It provides a good fit to $2011$ data, is consistent with the luminosities in $2004$ and $2008$, and reaches an agreement with the observed size.
We revise down the predicted radio and X-ray bow shock luminosities to be below the quiescent level of Sgr A*, which readily leads to non-detection in agreement to observations.
The magnetic energy dissipation in the cloud at the pericenter coupled with more powerful irradiation
may lead to an infrared source with an apparent magnitude $m_{L'}\approx13.0$. No shock into the cloud and no X-rays are expected from cloud squeezing by the ambient gas pressure.
Larger than previously estimated cloud mass $m_{\rm cloud}=(4-20)M_{\rm Earth}$ may produce a higher accretion rate and a brighter state of Sgr A* as the debris descend onto the BH.
\end{abstract}

\keywords{black hole physics --- Galaxy: center --- ISM: clouds --- magnetic fields --- radiation mechanisms: general --- radiative transfer}

\section{INTRODUCTION}\label{sec:intro}
The center of our Milky Way galaxy hosts a supermassive black hole (BH) Sgr A* with a mass $M_{\rm BH}=4.3\times10^{6}M_\odot$ located at a distance $d=8.3$~kpc \citep{Ghez2008,Gillessen:2009oo}.
The BH is primarily fed by hot stellar winds in the present epoch \citep{Cuadrawinds:2008,Shcherbakov:2010cond}, while clumps of cold gas provide an additional fuel source.
The accretion of such clumps may have been responsible for multiple Sgr A* outbursts observed as light echoes \citep{Clavel:2013qa,Czerny:2013aj}.
One of such objects first identified as a gas cloud G2 is observed on its way towards Sgr A* \citep{Gillessen:2012jq,Gillessen:2013lo}.
Its tail was hypothesized to be unrelated to the cloud, so that the compact cloud scenario is not excluded \citep{Phifer:2013ap}.
G2 is on a deeply plunging orbit with the pericenter distance in the range $r_p=1900-3100\rs$ \citep{Phifer:2013ap,Gillessen:2013pe},
where $\rs=2G M_{\rm BH}/c^2$ is the BH Schwarzschild radius. The center of mass (CM) of the cloud is predicted to pass the pericenter in $2013$ or $2014$.

The G2 cloud may have formed from the colliding stellar winds, which are subject to runaway cooling in the densest regions\citep{Cuadra:2005ot,Cuadrawinds:2008}.
The other cloud formation scenarios are a creation of a protoplanetary disk \citep{Murray-Clay:2012zb}, an encounter of a star with a stellar mass BH \citep{Miralda-Escude:2012al},
and a nova outburst \citep{Meyer:2012zf}. In all these cases the observed radiation comes from the gas photoionized and the dust heated by the intense starlight.
Alternatively, the object could host a central young star, which expels stellar wind \citep{Scoville:2013ql}.
The wind encounters ambient medium and produces a reverse shock. The observed emission then comes from the collisionally ionized shocked gas.

The object was extensively studied in the infrared (IR) band. The intrinsic luminosities of Brackett-$\gamma$ (Br$\gamma$), Paschen-$\alpha$ (Pa$\alpha$),
and Helium-$I$ (He$I$) lines were determined to yield $L({\rm Br}\gamma)=(1.06\pm0.32)10^{31}{\rm erg~s}^{-1}$ and the dereddened ratios
$L({\rm Pa}\alpha)/L({\rm Br}\gamma)=11.0\pm0.5$ and $L({\rm He}I)/L({\rm Br}\gamma)=0.8\pm0.3$ in the year $2011$ \citep{Gillessen:2013pe}.
The luminosities of these three lines are consistent with constants: of Br$\gamma$ from $2004$ till $2012$ and of the other two lines from $2008$ till $2011$.
A constant $L({\rm Br}\gamma)$ is not generally expected for a photoionized cloud \citep{Scoville:2013ql}.
The object was detected in $L'$ and $M$ bands with the dereddened absolute magnitudes $M_{L'}=-1.3\pm0.3$ and $M_M=-1.8\pm0.3$ in $2011$.
An upper limit with an apparent magnitude $m_{Ks}>(17-19)$~mag was measured in a $Ks$ band in $2011$ \citep{Gillessen:2012jq} and with $m_{Ks}\gtrsim20$~mag in $2012$ \citep{Phifer:2013ap}.
The latter limit is uncertain due to the source confusion \citep{Gillessen:2013pe,Eckart:2013nm}.

The observational manifestations of G2 passing close to Sgr A* were quantified. The bow shock during the pericenter passage accelerates a substantial amount of electrons.
Correspondent synchrotron radiation leads to an observable flux increase in the radio band \citep{Narayan:2012jl,Sadowski:2013mn}.
The radio flux predictions depend on the dynamics of the infall, in particular, on the cloud cross-section.
If G2 is a cloud, then it is tidally disrupted at the pericenter \citep{Gillessen:2012jq}.
After the disruption some debris fall back onto Sgr A*, which leads to a higher accretion rate and a higher Sgr A* luminosity \citep{Moscibrodzka:2012ax}.
The peak accretion rate and the duration of this state depend on the geometry and the cloud mass.
Thus, knowing the cloud properties is important for predicting the observational manifestations.
The cloud radius was observed to be $R_{\rm cloud}\sim15$~mas, which leads to the density $n_{\rm cloud}=2.6\times10^5{\rm cm}^{-3}$ and the mass $m_{\rm cloud}=3M_{\rm Earth}$ \citep{Gillessen:2012jq}.
These estimates are based on a simple photoionization of a spherical shape. Full radiative transfer calculations and fitting of the full IR dataset were not conducted.
Neither the consistency of the cloud hypothesis with the temporal behavior of the observed emission was quantitatively addressed.
The detailed emission diagnostics may either indicate that G2 is inconsistent with a gas cloud or confirm the cloud hypothesis and help to reliably determine the object properties.

In the present paper we perform such detailed analysis of G2 assuming a gas cloud paradigm.
We compute the incident continuum from a set of massive stars with known coordinates and luminosities.
We employ the radiative transfer code CLOUDY \citep{Ferland:1998by,Ferland:2013cp} to simulate the emission from the dust and the photoionized gas.
In Section~\ref{sec:models} we describe three models for the cloud shape: the spherical, tidally distorted, and magnetically arrested.
In Section~\ref{sec:emission} we present fitting of the simulated emission to the data.
We explore both the $10$~nm graphite dust grains and the full distribution of grain sizes incorporated into the interstellar medium (ISM) dust model.
We find good fits to $2011$ IR data and reproduce the temporal behavior of the IR emission with both the spherical and magnetically arrested cloud shapes.
In Section~\ref{sec:pericenter} we adopt the best-fitting magnetically arrested G2 model and estimate radiation during the pericenter passage.
The radio luminosity of the bow shock is revised down to below the quiescent level of Sgr A* owing to the smaller cloud cross-section.
In Section~\ref{sec:discussion} we discuss the results.
The observed IR magnitudes are commonly reported in Vega magnitude system $m\equiv m_{\rm Vega}$. The conversion between the AB system and Vega system
is frequency-dependent and is given by the formulas \citep{Bessell:1988mx,Bessell:1998ab,Tokunaga:2005ds,Blanton:2007et}
\begin{eqnarray}
&m&_{Ks,AB}-m_{Ks,\rm Vega}\approx1.86,\nonumber\\
&m&_{L',AB}-m_{L',\rm Vega}\approx2.97,\\
&m&_{M,AB}-m_{M,\rm Vega}\approx3.40,\nonumber
\end{eqnarray} while a single relation exists between the specific fluxes and the apparent AB magnitudes \citep{Oke:1974af}
\begin{equation}
m_{\nu,AB}=-2.5\log_{10}(F_\nu)-48.6.
\end{equation}
The extinction coefficients towards the Galactic Center are $A_{Ks}=2.42$, $A_{L'}=1.23$, and $A_M=1.07$ \citep{Fritz:2011jz}.

\section{DYNAMICAL MODELS OF THE G2 CLOUD}\label{sec:models}
\subsection{Spherical Cloud}
A sphere is the simplest shape. The spherical cloud is characterized by the radius $R_{\rm init}$ and the proton density $n_{\rm init}$.
The shape stays constant with time, which might not represent the physical behavior of a tenuous gas clump with weak self-gravity.
It was suggested that the additional gravity of the enclosed star may help to preserve the cloud shape \citep{Phifer:2013ap}.
A tenuous object is tidally disrupted, when it passes within the tidal radius
\begin{equation}\label{eq:tidal}
r_T=R_{\rm init}\left(\frac{M_{\rm BH}}{m_{\rm cloud}}\right)^{1/3}
\end{equation} from the BH.
If the central young T Tauri star is present, then its mass is around $1-2M_\odot$ \citep{Phifer:2013ap,Scoville:2013ql}.
Using the stellar mass $M_{\rm st}=2M_\odot$ in place of the cloud mass we obtain the tidal radius $r_T=2$~arcsec for the fiducial cloud radius $R=15$~mas \citep{Gillessen:2012jq}.
This tidal radius is comparable, but smaller than the cloud apocenter $r_{\rm apo}=1-2$~arcsec \citep{Gillessen:2013pe,Phifer:2013ap}.
Thus, the central star only helps to delay the tidal disruption of G2 on its way to Sgr A*.

\subsection{Tidally Distorted Cloud}
In the absence of the central star the tidal radius is $r_T=150$~arcsec for the fiducial cloud mass $m_{\rm cloud}=3M_{\rm Earth}$ \citep{Gillessen:2012jq}.
It is much larger than the apocenter radius, so that the cloud self-gravity can be neglected.
We consider the parts of the cloud to move independently in the gravitational field of Sgr A*.
On its way towards the center G2 is stretched along the direction of motion and compressed in the perpendicular direction just like a spherically symmetric accretion flow.
The tidally distorted cloud is characterized by the initial radius $R_{\rm init}$ and the initial density $n_{\rm init}$, similarly to the spherical model.
We denote by $r_{\rm init}$ the distance to Sgr A*, where the cloud is formed and where the spherical shape is assumed.
Let us derive the half-length $L$ of the cloud  based on the properties of its elliptical orbit.
In the following we consider the cloud to be sufficiently small, while all its parts are taken to move along the same trajectory.
Then the half-length is proportional to velocity of the cloud along its trajectory $L\propto v$.
The angular momentum conservation of the cloud is
\begin{equation}
v_\phi r=\rm const,
\end{equation} where $v_\phi$ is the angular velocity at the distance $r$ from Sgr A*.
The energy conservation reads
\begin{equation}
\frac{v_\phi^2+v_r^2}{2}-\frac{G M_{BH}}{r}=\rm const,
\end{equation} where $v_r$ is the radial velocity.
The velocity along the trajectory $v=\sqrt{v_r^2+v_\phi^2}$ is then
\begin{equation}
v=\sqrt{\frac{2G M_{BH}(r_{\rm apo}+r_p-r)}{r(r_{\rm apo}+r_p)}}.
\end{equation}
The resultant half-length is
\begin{equation}\label{eq:length}
L=R_{\rm init}\left(\frac{r_{\rm init}}{r}\right)^{1/2}\left(\frac{r_{\rm apo}+r_p-r}{r_{\rm apo}+r_p-r_{\rm init}}\right)^{1/2}.
\end{equation}
This formula explains dramatic elongation of simulated clouds forming at the apocenter \citep{Burkert:2012ca,Schartmann:2012ma}.
If the cloud starts far from the apocenter, then the expression simplifies to
\begin{equation}\label{eq:lengthapprox}
L\approx R_{\rm init}\left(\frac{r_{\rm init}}{r}\right)^{1/2}.
\end{equation}
The perpendicular size $\rho$ depends on the initial velocity distribution in the cloud, but if G2 behaves like a converging flow inward of the apocenter, then
$\rho$ is proportional to the distance
\begin{equation}\label{eq:tidsize}
\rho=R_{\rm init}\frac{r}{r_{\rm init}}.
\end{equation} The converging flow behavior is natural for the cloud formation via the cooling instability. In contrast, for their calculations \citet{Gillessen:2012jq} assumed
isotropic Gaussian initial velocity distribution of cloud particles with high dispersion $\sigma=120{\rm km~s}^{-1}$.
The resultant cloud density is
\begin{equation}\label{eq:tiddense}
n=n_{\rm init}\left(\frac{r_{\rm init}^2 L}{r^2R_{\rm init}}\right)\approx n_{\rm init}\left(\frac{r_{\rm init}}{r}\right)^{3/2}.
\end{equation}

The cloud may have formed near the apocenter, where it spends most of its time along the orbit.
However, according to formula~(\ref{eq:length}) such a cloud becomes very elongated, which is inconsistent with the observed size.
The latest estimates of the G2 orbit give the apocenter radius \citep{Gillessen:2013pe,Phifer:2013ap}
\begin{equation}
r_{\rm apo}\approx2{\rm arcsec}
\end{equation} and the pericenter radius
\begin{equation}
r_{\rm p}\approx0.02{\rm arcsec},
\end{equation}
while we take the formation distance to be
\begin{equation}
r_{\rm init}=1{\rm arcsec}
\end{equation} inward of the apocenter.
The described tidally distorted shape is more suitable, than the spherical shape, for the cloud without the central source, but it assumes zero magnetic field.

\subsection{Magnetically Arrested Cloud}
Strong magnetic field alters the cloud shape. We denote by $\sigma$ the ratio of the magnetic field energy density to the gas energy density in the cloud at the place of formation, such that
\begin{equation}
\frac{B_{\rm init}^2}{8\pi}=\sigma 3k_B n_{\rm init} T_{\rm cloud},
\end{equation} where the cloud temperature is approximately $T_{\rm cloud}\sim8\times10^3$~K \citep{Gillessen:2012jq}.
The magnetic flux conservation along the direction of motion
\begin{equation}\label{eq:Bfield}
B_{\rm init}R_{\rm init}^2=B_{||} \rho^2
\end{equation} leads to a substantial growth of the B-field and the correspondent magnetic forces as in
the spherical magnetized accretion flow \citep{Shvartsman:1971,Shcherbakov:2008io}.
Two dominant forces acting on the cloud perpendicular to its orbital plane are the gravitational force
\begin{equation}\label{eq:Fg}
F_{\rm g}=\frac{G M_{\rm BH}m_{\rm cloud}\rho}{r^3},
\end{equation} where the cloud mass is $m_{\rm cloud}=(2R_{\rm init})^3 m_p n_{\rm init}$, and the magnetic force
\begin{equation}\label{eq:Fmagn}
F_{\rm magn}=\frac{B_{||}^2}{8\pi}4L\rho.
\end{equation}
The hot ambient gas force $F_{\rm out}$ and the cloud gas force $F_{\rm in}$ can be comparable to the magnetic force at the formation distance,
but are sub-dominant inward as the magnetization of the cloud grows.

The parallel force balance is dominated by the rapidly growing gravitational force $F_{\rm g,||}\propto L/r^3\propto r^{-7/2}$, so that the cloud is stretched in the parallel direction
according to the equation~(\ref{eq:length}) as in the tidally distorted model.
Solving the perpendicular force balance
\begin{equation}\label{eq:fbalance}
F_{\rm g}=F_{\rm magn},
\end{equation} we find the perpendicular radius
\begin{equation}\label{eq:rho}
\rho=38{\rm mas}~\left(\frac{r}{\rm arcsec}\right)^{5/8}\left(\frac{r_{\rm init}}{\rm arcsec}\right)^{1/8}\sqrt{\frac{R_{\rm init}}{0.1{\rm arcsec}}}\sigma^{1/4},
\end{equation} which depends weakly on both the initial cloud magnetization $\sigma$ and the formation distance.
The distance from Sgr A*, where the cloud becomes magnetically arrested is
\begin{equation}
r_{\rm crit}=78{\rm mas}~\left(\frac{r_{\rm init}}{\rm arcsec}\right)^3\left(\frac{R_{\rm init}}{0.1\rm arcsec}\right)^{-4/3}\sigma^{2/3}.
\end{equation} There the perpendicular cloud size switches from $\rho\propto r$ to $\rho\propto r^{5/8}$ behavior.
The critical distance $r_{\rm crit}$ equals the formation distance $r_{\rm init}=1$~arcsec for the initial magnetization
\begin{equation}\label{eq:Rinitx}
\sigma_x=\left(\frac{R_{\rm init}}{15{\rm mas}}\right)^2.
\end{equation} The density is approximately
\begin{equation}\label{eq:magdense}
n\approx n_{\rm init}\left(\frac{r}{r_{\rm init}}\right)^{-3/4}
\end{equation} of the cloud, which forms magnetically arrested.
The column density of such a cloud is practically constant with distance to Sgr A* as $n \rho\propto r^{1/8}$.
We define the free-fall time at the distance $r$ as
\begin{equation}
t_{\rm ff}=\frac{r}{v_K},~{\rm where}~v_K=\sqrt{\frac{G M_{\rm BH}}{r}}
\end{equation} is the Keplerian velocity.
The efficient magnetic energy dissipation occurs on the Alfven timescale
\begin{equation}
t_{\rm A}=\frac{\rho}{v_{\rm A}},~{\rm where}~v_{\rm A}=\frac{B_{||}}{\sqrt{4\pi n m_p}}
\end{equation} is the Alfven speed.
The ratio of these timescales is always
\begin{equation}
\frac{t_{\rm A}}{t_{\rm ff}}=\frac12
\end{equation} in the magnetically arrested regime.
If efficient energy dissipation happens on the dynamical time, then the cloud shape approaches the tidally distorted shape.
However, the continuous magnetic field dissipation during the formation and the initial motion of the cloud is likely to leave the large scale ordered magnetic field
or the highly helical magnetic field \citep{Biskamp:2003zb,Shcherbakov:2008io}. The resultant inefficient dissipation likely occurs on the large timescale
\begin{equation}
t_{\rm diss}\gg t_{\rm A}.
\end{equation} We adopt the latter case of weak dissipation and neglect the influence of the finite $t_{\rm diss}$ on the cloud shape.
We fit the $2011$ observational data and explore the temporal behavior of the models for all three presented shapes.

\section{EMISSION LINES AND DUST DIAGNOSTICS}\label{sec:emission}
\subsection{Incident Radiation}
The incident radiation flux and spectrum need to be reliably determined to model the cloud.
We quantify the incident radiation starting from the properties of the individual bright stars in the Galactic Center region.
We take the positions and the velocities of the bright stars from \citet{Paumard2006,LuGhez:2009} and the stellar temperatures and luminosities from \citet{Martins:2007}.
Following \citet{Cuadrawinds:2008} we correct the sample of Wolf-Rayet stars for completeness.
The incident flux emitted by the bright stars is dominated at Sgr A* by IRS16NW, IRS16C, and IRS16SW.
The vertical offsets $z$ of IRS16NW and IRS16C are not known, but their $z$ velocities are much larger than that of IRS16SW \citep{LuGhez:2009}.
Then we assume IRS16NW and IRS16C to have zero vertical offsets $z=0$. The rest of the bright stars contribute about $1/3$ of the total flux from these three.
The IRS16 stars are more than $1$~arcsec away from Sgr A*. The closer in, but less luminous, S stars might substantially contribute.
We explicitly include S0-2 star into the calculations as one of the most luminous and the closest to Sgr A* S stars \citep{Martinss2:2008,Gillessen:2009s2}.
Dimmer S stars contribute relatively little to the incident ionizing flux, when G2 is far from the pericenter.
Their initial mass function (IMF) has the slope $\Gamma=-2.15\pm0.3$ \citep{Do:2013kl}. However, the dependence of their bolometric luminosity on mass
is very steep $L_\star\propto M_\star^{3.5}$ at $M_\star\lesssim10M_\odot$ (e.g. \citealt{Salaris:2005al}) and the dependence
of the luminosity above the hydrogen ionization threshold is even steeper.
Heavy stars with the masses $M_\star>10M_\odot$ and the top-heavy IMF \citep{Bartko:2010zs} are included into the calculation individually.

The incident fluxes at the locations of G2 and Sgr A* are presented in Table~\ref{tab:obs}. The positions of G2 in the picture plane are taken from \citet{Phifer:2013ap},
while the inclination angle $i=118^\circ$ and the longitude of periastron $\omega=97^\circ$ are taken from \citet{Gillessen:2013pe}.
The S0-2 star was far from its pericenter in the years of observations.
The incident flux changed by only $30\%$ between $2004$ and $2011$ despite the G2 cloud moved substantially.
The total flux
\begin{equation}
F_{\rm tot}=5.2\times10^4{\rm erg~s}^{-1}{\rm cm}^{-2}
\end{equation}
and the photon energy density
\begin{equation}
U_{\rm ph}=1.7\times10^{-6}{\rm erg~cm}^{-3}
\end{equation} at Sgr A* in $2011$ are an order of magnitude larger than the estimates in \citet{Krabbe:1991xf,Quataert:2005no},
but are much smaller than the flux and the energy density at Sgr A*, when S0-2 passes through its pericenter \citep{Nayakshin:2005co}.
\begin{table*}
\caption{Distances and Incident Fluxes at the Locations of G2 and Sgr A* in Different Years}\label{tab:obs}
\begin{tabular}{ | p{50mm} | p{22mm}| p{22mm} | p{22mm} | p{22mm} | p{22mm} | }
\tableline\tableline
  Quantity  & In $2004$ at G2 & In $2008$ at G2 & In $2011$ at G2 & At G2 pericenter & In $2011$ at Sgr A* \\ \tableline
  Distance from G2 to Sgr A*, arcsec \tablenotemark{a}& 0.59 & 0.43 & 0.30 & 0.020 & \nodata \\\tableline
  Total flux\tablenotemark{b,c}, $10^4{\rm erg~s}^{-1}{\rm cm}^{-2}$ & $3.0~(75\%)$ & $3.5~(87\%)$ & $4.0~(100\%)$ & $5.7~(142\%)$& $5.2~(129\%)$  \\\tableline
  S0-2 flux\tablenotemark{b},$10^4{\rm erg~s}^{-1}{\rm cm}^{-2}$ & $0.26~(31\%)$ & $0.47~(57\%)$ & $0.84~(100\%)$ & $2.6~(306\%)$ & $2.0~(241\%)$  \\\tableline
  Total of IRS16NW, IRS16C, and IRS16SW fluxes\tablenotemark{b},$10^4{\rm erg~s}^{-1}{\rm cm}^{-2}$ & $2.8~(86\%)$ & $3.1~(95\%)$ & $3.2~(100\%)$ & $3.2~(100\%)$ & $3.2~(100\%)$ \\\tableline
  S0-2 contribution to flux & $8.4\%$ & $13\%$ & $21\%$ & $44\%$ & $39\%$  \\\tableline
\end{tabular}
\tablenotetext{1}{For the orbital inclination angle $i=118^\circ$ and the longitude of periastron $\omega=97^\circ$.}
\tablenotetext{2}{Fluxes relative to $2011$ are shown in parentheses.}
\tablenotetext{3}{Computed as the sum of the S0-2 flux and $4/3$ of the total flux from IRS16NW, IRS16C, and IRS16SW.}
\end{table*}
\subsection{Emission of the Cloud}
The radiative transfer through the mixture of the ionized gas and the dust is performed with the version 13 of CLOUDY code \citep{Ferland:2013cp}.
We assume the cloud is in ionization/recombination equilibrium.
The recombination time is
\begin{equation}\label{eq:trec}
t_{\rm rec}=\frac{1}{\alpha_B n_{\rm cloud}}\approx1.0\left(\frac{10^5{\rm cm}^{-3}}{n_{\rm cloud}}\right){\rm yrs}
\end{equation}
for the case B \citep{Draine:2011ism} with recombination coefficient $\alpha_B\approx3\times10^{-13}{\rm cm}^3{\rm s}^{-1}$ \citep{Osterbrock:2006ag} at $T=8\times10^3$~K.
This timescale is smaller than the time difference between the years of interest 2004, 2008, and 2011, unless the cloud of low density $n_{\rm cloud}=3\times10^4{\rm cm}^{-3}$ is considered.
Even then the recombination time is shorter than the characteristic radiative timescale of the system, as the ionizing flux varies only weakly with time.

We effectively consider the stretched cloud to be rectangular and irradiated perpendicular to its longer side from one direction.
We compute models with the graphite dust with $10$~nm grains and with the ISM dust, which includes a range of grain sizes \citep{Mathis:1977de}.
The former choice is motivated by the high inferred dust temperature, which can only be achieved by the very small grains \citep{Draine:2011ism,Gillessen:2012jq}.
We consider the spherical, tidally distorted, and magnetically arrested cloud shapes.
We vary the dust abundance $A$ relative to the ISM \citep{Bohlin:1978da} and the cloud radius $R_{\rm init}$ fitting the $2011$ IR data for a set of the initial cloud densities $n_{\rm init}$.
We fit the Br$\gamma$ luminosity, the ratio $L({\rm He}I)/L({\rm Br}\gamma)$, and the luminosities in $M$ and $L'$ bands.
We do not fit the ratio of the Pa$\alpha$ luminosity to the Br$\gamma$ luminosity, as it is practically constant
\begin{equation}
L({\rm Pa}\alpha)/L({\rm Br}\gamma)=11.7-11.8
\end{equation} for all computed models. The values $11.7-11.8$ are within $2\sigma$ from the observed ratio.
We consider the transmitted continuum, but check that the reflected continuum is consistent with it to within $10\%$ in each IR band of interest for all computed models.
The He$I$ line shows the largest optical depth $\tau$ among the lines of interest, but it is still optically thin with $\tau=0.03-0.1$.

\begin{figure}[htbp]
    \centering\plotone{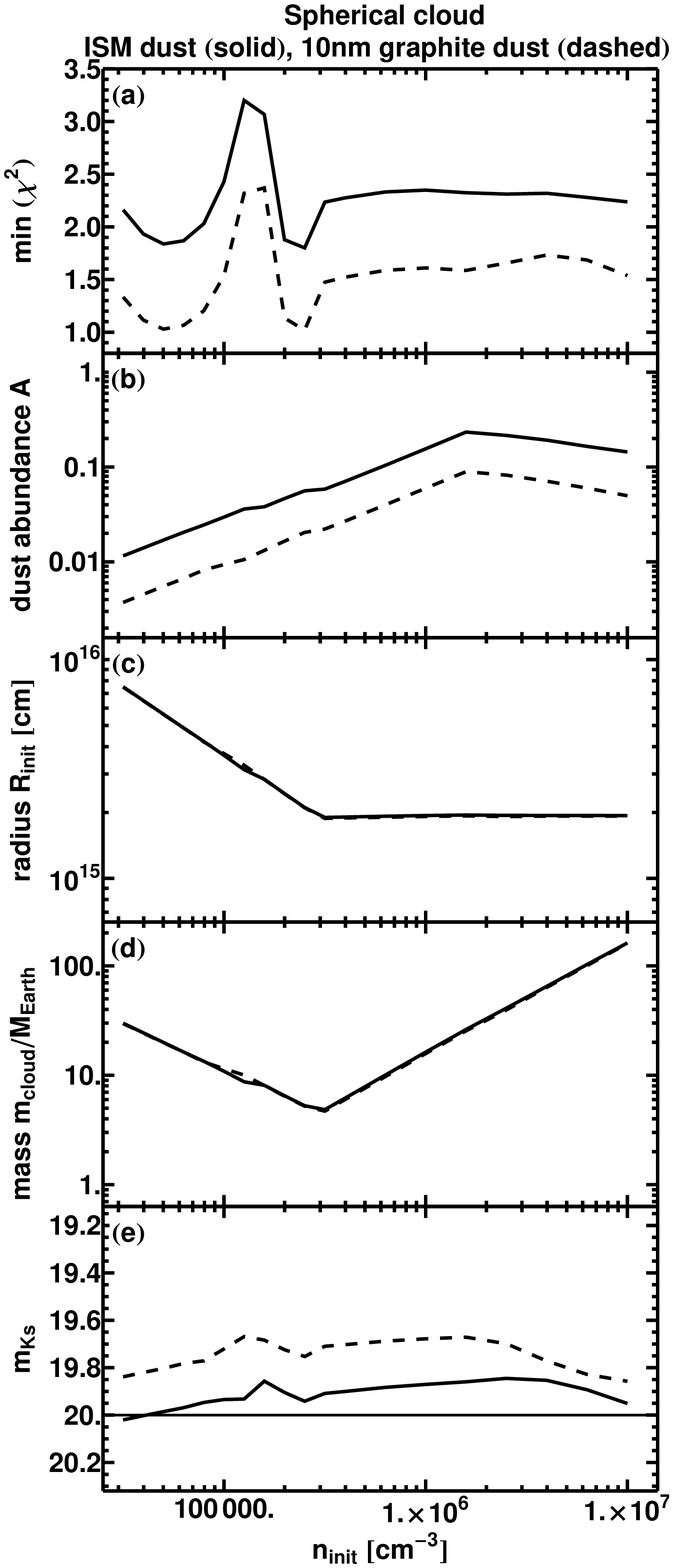}
    \caption{Properties of the best-fitting spherical clouds for the different densities $n_{\rm init}$: the minimum $\chi^2$ for fitting $L({\rm Br}\gamma)$,
    $L({\rm He}I)/L({\rm Br}\gamma)$, $M_{L'}$, and $M_M$ (panel a), the relative amount of dust $A$ (panel b), the initial cloud radius $R_{\rm init}$ (panel c),
    the cloud mass measured in the masses of the Earth $m_{\rm cloud}/M_{\rm Earth}$ (panel d), and the simulated apparent $Ks$ magnitude $m_{Ks}$ (panel e).
    Shown are the models with the ISM dust (solid) and the $10$~nm graphite dust (dashed). The tentative observational upper limit in $Ks$ band is shown as a horizontal line in the bottom panel.}
    \label{fig:spherical_fit}
\end{figure}

The properties of the spherical models, which provide the best fits to the data at each density,
are shown in Figure~\ref{fig:spherical_fit} for the ISM dust (solid lines) and the $10$~nm graphite dust (dashed lines).
The models with the graphite dust fit the data very well for a wide range of densities $n_{\rm init}=3\times10^4-10^7{\rm cm}^{-3}$.
The exception is a narrow region of a poor fit around $n_{\rm init}=1.5\times10^5{\rm cm}^{-3}$, where the simulated ratio $L({\rm He}I)/L({\rm Br}\gamma)$ is too high.
The models with the ISM dust behave similarly to the models with fine graphite dust, though the fit to the data is systematically worse, since ISM dust emission is too red.
The best-fitting cloud mass is around $m_{\rm cloud}=5M_{\rm Earth}$, but the high-density models with large cloud masses $m_{\rm cloud}=(10-100)M_{\rm Earth}$ do fit the data well.
The observed $Ks$ magnitude $m_{Ks}=19.8-20.0$ (absorbed by the ISM) is consistent with the non-detection.
The best-fitting initial cloud radius $R_{\rm init}=2\times10^{15}{\rm cm}=16$~mas equals the observed spatial extent of the cloud \citep{Gillessen:2012jq}.
while the low-density cloud models are inconsistent with the observed size. The minimum $\chi^2$ (top panel) has two minima as a function of density, but the cloud models given by
the low density minima overpredict the cloud spatial extent. Thus, here and below we concentrate on the models given by the high density minima.
The models consistent with the observations have small recombination time $t_{\rm rec}<0.5$~yrs,
which validates our recombination equilibrium approximation.

\begin{figure}[htbp]
    \centering\plotone{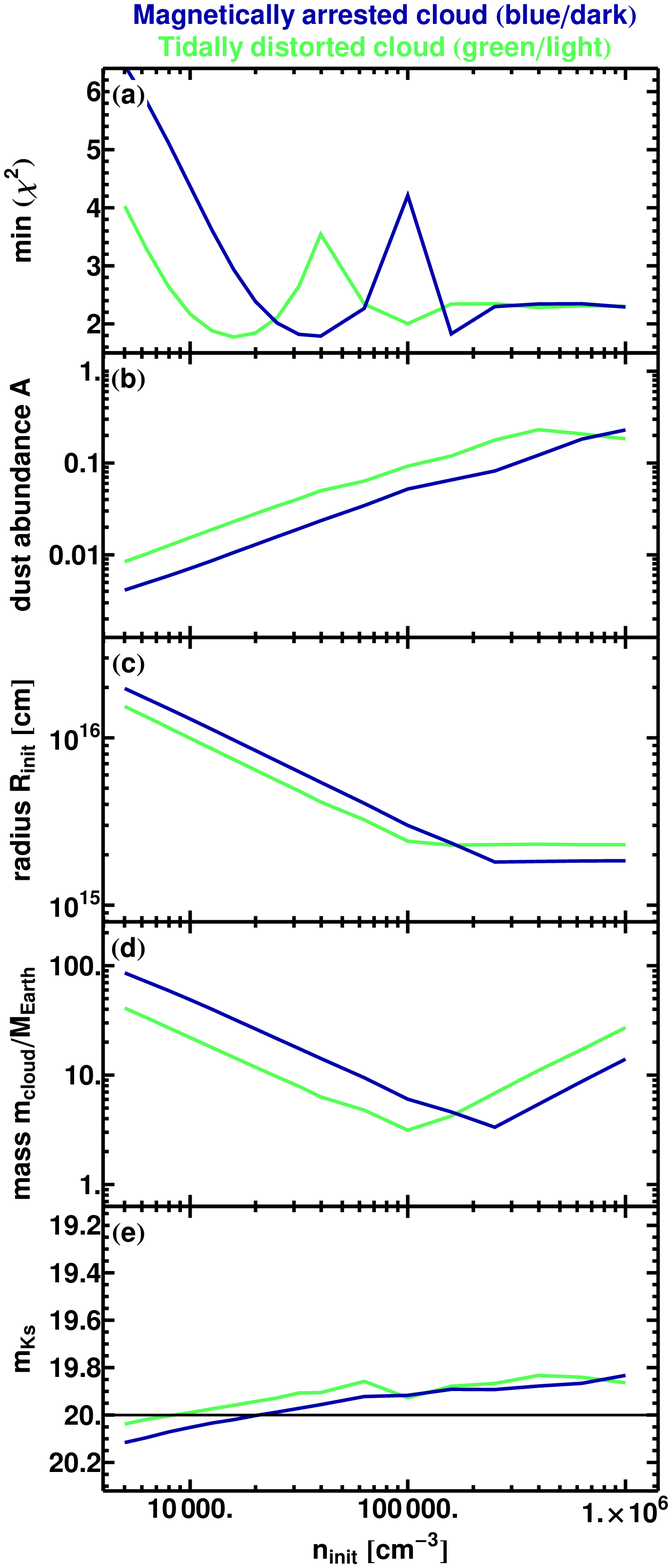}
    \caption{Properties of the best-fitting magnetically arrested (blue/dark lines) and tidally distorted (green/light lines) cloud models
    with the ISM dust for the different initial densities $n_{\rm init}$: the minimum $\chi^2$ for fitting $L({\rm Br}\gamma)$, $L({\rm He}I)/L({\rm Br}\gamma)$,
    $M_{L'}$, and $M_M$ (panel a), the relative amount of dust $A$ (panel b), the initial cloud radius $R_{\rm init}$ (panel c), the cloud mass measured in the masses of the Earth
    $m_{\rm cloud}/M_{\rm Earth}$ (panel d), and the simulated apparent $Ks$ magnitude $m_{Ks}$ (panel e). The tentative observational upper limit in $Ks$ band
    is shown as a horizontal line in the bottom panel.}
    \label{fig:tidal_fit}
\end{figure}

The properties of the magnetically arrested models (blue/dark lines) and the tidally distorted models (green/light lines), which provide the best fit to the data at each density,
are shown in Figure~\ref{fig:tidal_fit} for the ISM dust.
The best-fitting initial densities for these shapes are $n_{\rm init}=1.8\times10^5{\rm cm}^{-3}$ and $n_{\rm init}=1.0\times10^5{\rm cm}^{-3}$, respectively.
These densities correspond to about the same cloud density $n=(3-5)\times10^5{\rm cm}^{-3}$ in $2011$.
The best-fitting masses of the cloud $m_{\rm cloud}$, the relative dust abundances $A$, and the simulated apparent $Ks$ magnitudes practically coincide with the values for the best-fitting spherical model.
The initial cloud radii $R_{\rm init}\lesssim3\times10^{15}$~cm are consistent with the observations due to the shrinking of the cloud in the perpendicular direction.
We discuss the consistency with the size observations in more detail below for the best-fitting model.

\begin{figure*}[htbp]
    \centering\plotone{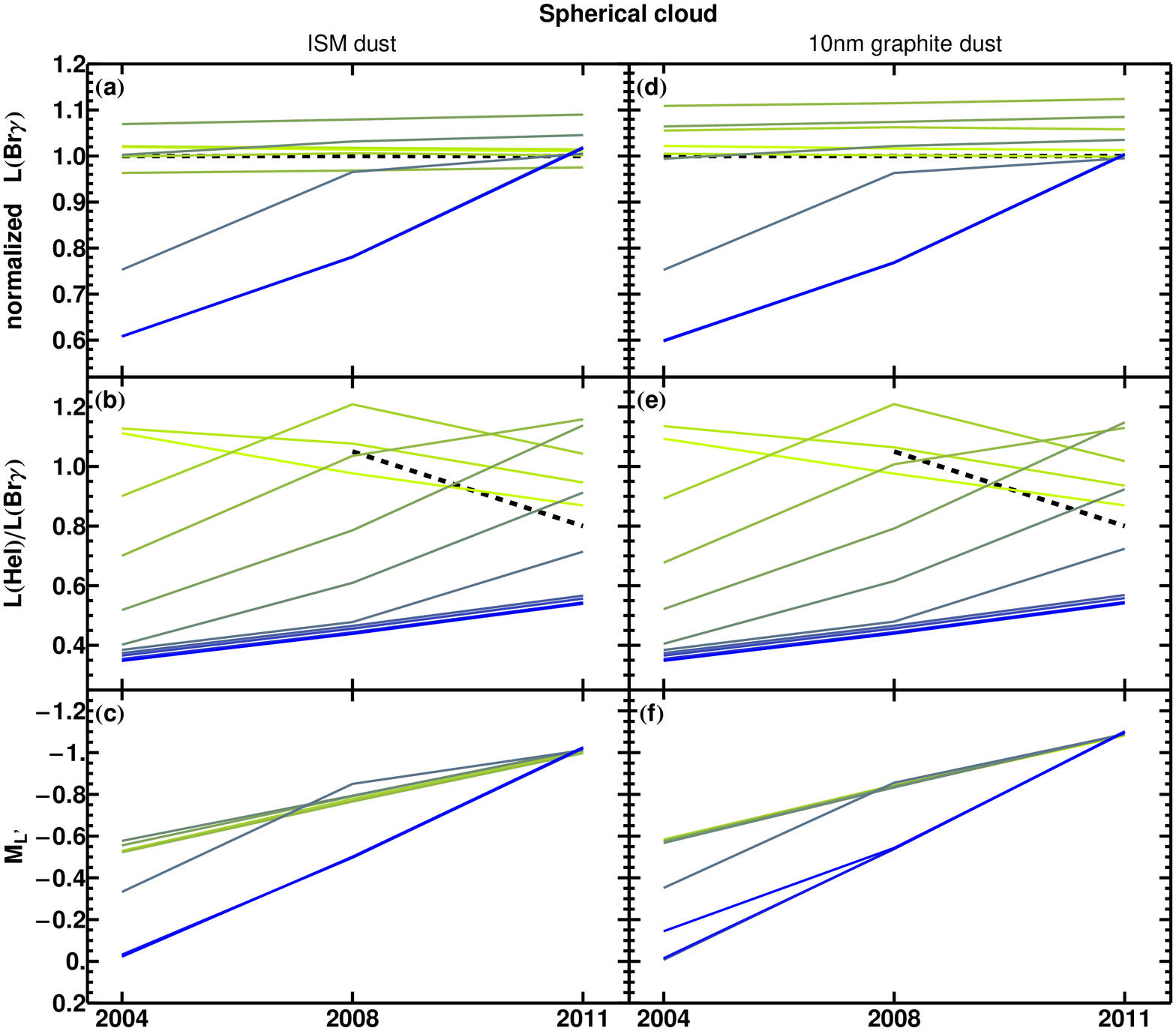}
    \caption{Temporal dependence of quantities in the best-fitting spherical models for a range of the initial cloud densities $n_{\rm init}$:
    the Br$\gamma$ luminosity normalized to the observed line luminosity (top row), the ratio $L({\rm He}I)/L({\rm Br}\gamma)$ (middle row), and the absolute magnitude $M_{L'}$ (bottom row).
    The models with the initial densities $\log (n_{\rm init}, {\rm cm}^{-3})=4.8, 4.9, 5.0, 5.1, 5.2, 5.3$ correspond to the optically thin group (light/yellow lines),
    while the models with the initial densities $\log (n_{\rm init}, {\rm cm}^{-3})=5.4, 5.5, 5.6, 5.8, 6.0, 6.2$ correspond to the optically thick group (dark/blue lines).
    The left column shows the results for the ISM dust and the right column shows the results for the $10$~nm graphite dust.  The thick dashed lines show the observations.}
    \label{fig:spherical_time}
\end{figure*}

The critical test for any cloud model is the ability to reproduce the IR luminosities obtained in different years.
In Figure~\ref{fig:spherical_time} we present the simulated normalized Br$\gamma$ luminosity, the ratio $L({\rm He}I)/L({\rm Br}\gamma)$, and the absolute magnitude $M_{L'}$
for the spherical clouds in the years $2004$, $2008$, and $2011$. The models tested are the ones, which provide the best fits to the $2011$ data for each initial density.
We find two large classes of the outcomes different by the optical depth to the irradiating ionizing continuum.
The optically thin models with low $n_{\rm init}$ do not substantially attenuate the ionizing flux and maintain the high gas temperature $T_{\rm cloud}=(0.7-1.2)\times10^4$~K throughout the cloud.
The optically thick models with high $n_{\rm init}$ absorb most of the ionizing radiation, so that their temperature drops to $T_{\rm cloud}\lesssim5\times10^3$~K on the far side of the cloud.
The latter models emit most of their IR flux near the irradiated side of the cloud.
The optically thin spherical clouds exhibit practically constant with time $L({\rm Br}\gamma)$.
Since the cooling function depends strongly on the temperature $\Lambda(T)\propto T^\delta$ with $\delta\sim5$ at $T\sim10^4$~K,
then the increase by $30\%$ of the irradiating flux between the years $2004$ and $2011$ leads to only a $6\%$ temperature rise.
The line emissivity \citep{Draine:2011ism}
\begin{equation}
\alpha({\rm Br}\gamma)\propto T^{-1}n^2
\end{equation} then decreases by $6\%$, which explains the trend at the lowest simulated densities.
The increase of $L({\rm Br}\gamma)$ in the optically thick group simply follows the rise of the irradiating flux, since both the density and the cross-section of the spherical cloud remain constant.
The ratio $L({\rm He}I)/L({\rm Br}\gamma)$ slightly decreases in the optically thin models between $2008$ and $2011$ consistently with the observations.
The flux emitted by the dust always positively correlates with the incident flux and the source becomes brighter with time in $L'$ band.
The temporal dependence of the $M$-band absolute magnitude directly follows the dependence of the $L'$ magnitude shown in the bottom panel.
The best-fitting model lies within the optically thin group and thus not only explains the spectrum in $2011$, but also reproduces the observations in the earlier years.
The models with the ISM dust and the graphite dust behave similarly.

\begin{figure*}[htbp]
    \centering\plotone{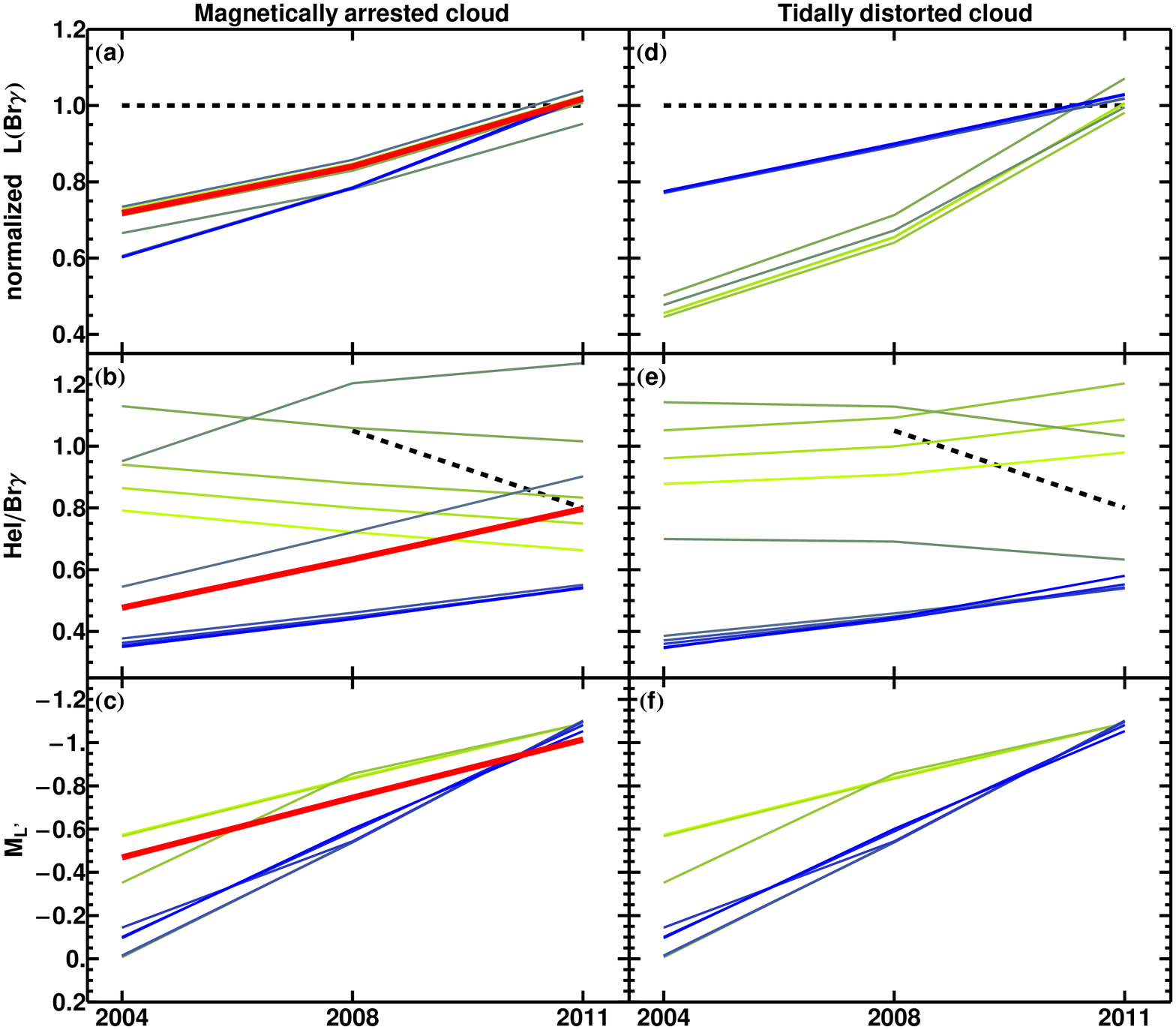}
    \caption{Temporal dependence of quantities in the best-fitting magnetically arrested (left column) and tidally distorted (right column) models for a range of the initial cloud densities $n_{\rm init}$:
the Br$\gamma$ luminosity normalized to the observed line luminosity (top row), the ratio $L({\rm He}I)/L({\rm Br}\gamma)$ (middle row), and the absolute magnitude $M_{L'}$ (bottom row).
    The models with the initial densities $\log (n_{\rm init}, {\rm cm}^{-3})=4.4, 4.5, 4.6, 4.8, 5.0$ correspond to the optically thin group (light/yellow lines),
    while the models with the initial densities $\log (n_{\rm init}, {\rm cm}^{-3})=5.2, 5.4, 5.6, 5.8, 6.0$ correspond to the optically thick group (dark/blue lines).
        The thick dashed lines show the observations. The thick solid/red lines in the left column show the best-fitting magnetically arrested model.}
  \label{fig:tidal_time}
\end{figure*}

In Figure~\ref{fig:tidal_time} we present the temporal behavior of the tidally distorted and the magnetically arrested models with the ISM dust.
The models tested are the ones, which provide the best fits to the $2011$ data for each initial density.
The tidally distorted shapes show distinct behaviors for the optically thin and the optically thick groups.
The Br$\gamma$ luminosity rises steeply in the optically thin group. This is due to a large density increase with time and is inconsistent with the observations.
The ratio $L({\rm He}I)/L({\rm Br}\gamma)$ increases with time in a tidally distorted model.
In turn, the optically thin and the optically thick magnetically arrested models behave similarly, owing to the less dramatic temporal changes of density.
The rise of the Br$\gamma$ luminosity by $1.3$ from $2004$ to $2011$ is marginally consistent with the observations of this line \citep{Gillessen:2013pe}.
We conduct the linear regression over the observed values of $L({\rm Br}\gamma)$ and find that the variations by $30\%$ between $2004$ and $2011$ are near the boundary of the $90\%$ confidence interval.
The optically thin magnetically arrested models have a small negative temporal slope of the $L({\rm He}I)/L({\rm Br}\gamma)$ ratio in agreement to observations.

\subsection{Best-fitting Magnetically Arrested Model}
The best-fitting magnetically arrested cloud model is marginally optically thick and produces relatively low He$I$ luminosity in $2008$, but the models with slightly lower density readily reproduce
He$I$ to Br$\gamma$ ratio. This model has the initial density, the initial radius, and the relative dust abundance of
\begin{equation}
n_{\rm init}=1.8\times10^5{\rm cm}^{-3},~R_{\rm init}=2.2\times10^{15}{\rm cm}=17 {\rm mas},~\text{and} A=0.072,
\end{equation} respectively, and reaches $\chi^2=1.70$. This relative dust abundance corresponds to the dust-to-gas ratio $4.6\times10^{-4}$ by mass,
while the total mass of such cloud is $m_{\rm cloud}=4.1M_{\rm Earth}$.
The cloud with the initial size $R_{\rm init}=17{\rm mas}$ shrinks in the perpendicular direction to $\rho=10$~mas and $\rho=8$~mas in $2008$ and $2011$, respectively, which is consistent with
the direct size measurements \citep{Gillessen:2012jq}.
The radial stretching to the half-length $L=41$~mas produces the projected size $R=20$~mas in $2011$ for the inclination angle $i=118^\circ$ and
the longitude of periastron $\omega=97^\circ$. This spatial extent is consistent with the observations. Thus, radial stretching dominates the observed size.

\begin{figure*}[htbp]
    \centering\plotone{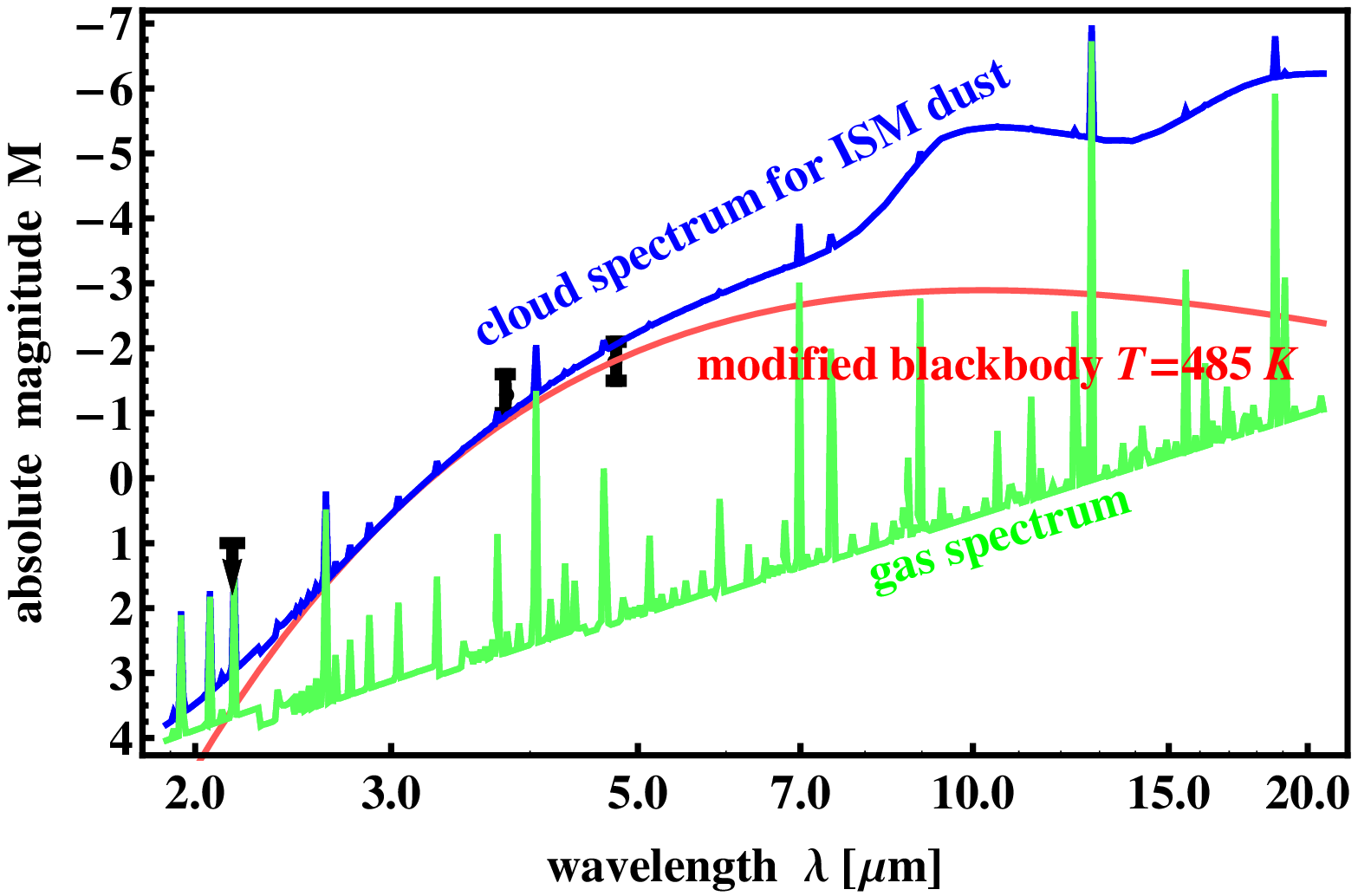}
    \caption{IR spectrum of the best-fitting magnetically arrested model in $2011$ (blue/dark line), modified blackbody contribution of the smallest dust grains (red/light smooth line),
    gas contribution (green/light edgy line), and observations reported in \citet{Gillessen:2012jq} (error bars and upper limit in black/dark).}
    \label{fig:spectrum}
\end{figure*}

The velocity spread between two ends of the cloud is
\begin{equation}\label{eq:dv}
\Delta v\approx\sqrt{\frac{2G M_{\rm BH}}{r^3}}L \sin i\sin\omega,
\end{equation} when the object is on a highly eccentric orbit far from the pericenter.
The values $\Delta v=205{\rm km~s}^{-1}$ in $2011$ and $\Delta v=100{\rm km~s}^{-1}$ in $2008$ for the chosen orbital parameters are lower
than the intrinsic integrated full-width at half-maximum velocity $\Delta v_{\rm obs}=350\pm40{\rm km~s}^{-1}$ in $2011$ and $\Delta v_{\rm obs}=210\pm24{\rm km~s}^{-1}$ in $2008$
reported in \citet{Gillessen:2012jq} for the head of the cloud.
This indicates that the cloud is either initially bigger than in the best-fitting model or that the cloud started closer to the apocenter.
The spectrum of this model in $2011$ is shown in Figure~\ref{fig:spectrum} (blue line).
The modified blackbody spectrum (smooth red line)  represents the contribution of the smallest dust grains with sizes $6-13$~nm, which have the temperature $T_{\rm dust}=485$~K,
The larger grains, which have the lower temperatures $T_{\rm dust}=300-450$~K, contribute substantially to the observed emission in the $M$ and $L'$ bands.
The emissivity of the smallest particles $\epsilon_\nu$ is given by the modified blackbody spectrum \citep{Draine:2011ism}
\begin{equation}
\epsilon_\nu\propto \nu^2 B_\nu\propto\nu^5\left(\exp\left[\frac{h \nu}{k_B T_{\rm dust}}\right]-1\right)^{-1},
\end{equation} where $B_\nu$ is the blackbody spectrum. The emission in $Ks$ band is dominated by gas, whose contribution is shown in the edgy green line in Figure~\ref{fig:spectrum}.
 The conversion of AB magnitudes into Vega magnitudes to produce Figure~\ref{fig:spectrum} is performed by
assuming Vega IR spectrum to be a perfect blackbody with temperature $T=9450$~K as done in \citet{Gillessen:2012jq}.

The cloud is magnetically arrested at a distance $r_{\rm init}=1$~arcsec, when its magnetization is $\sigma\sim2$ according to the formula~(\ref{eq:Rinitx}).
The ambient hot gas density and temperature are approximately \citep{Baganoff:2003,Shcherbakov:2010cond}
\begin{equation}\label{eq:gasdens}
n_{\rm gas}=130\left(\frac{r}{\rm arcsec}\right)^{-1}{\rm cm}^{-3}
\end{equation} and
\begin{equation}\label{eq:gastemp}
 T_{\rm gas}=3\times10^7\left(\frac{r}{\rm arcsec}\right)^{-1}{\rm K}.
\end{equation}
The gas pressure of the best-fitting magnetically arrested model is about $40\%$ of the ambient gas pressure at the formation $n_{\rm init} T_{\rm cloud}\sim 0.4n_{\rm gas} T_{\rm gas}$.
The cloud magnetic force $F_{\rm magn}$ is comparable to the ambient gas force $F_{\rm out}$, thus cloud is initially close to pressure equilibrium.
 However, the magnetic force grows faster inwards, which justifies neglecting the ambient gas in calculation of the shape. Let us estimate the radiation from such cloud, when it passes through the pericenter.

\section{RADIATION DURING PERICENTER PASSAGE}\label{sec:pericenter}
\subsection{Bow Shock Radio, IR, and X-ray Emission}
As the cloud passes through the pericenter, it creates a bow shock with a Mach number $\mathcal{M}\approx2$, which accelerates the electrons \citep{Narayan:2012jl,Sadowski:2013mn}.
The accelerated electrons radiate synchrotron emission, most notably in the radio band.
The particle distribution was formerly estimated to be
\begin{equation}\label{eq:distRamesh}
\frac{dN}{d\gamma}\approx 2\times10^{49}\gamma^{-2.2}, \quad \gamma\ge2
\end{equation} for the fiducial efficiency $5\%$ and the cloud perpendicular radius $\rho_{\rm Ram}=10^{15}$~cm at the pericenter.
The ambient hot gas density is $n\approx4\times10^3{\rm cm}^{-3}$ and the temperature is $T\approx10^9$~K at the adopted pericenter radius $r_p=2000\rs$ 
\citep{Yuan:2003sg,Xu:2006dw,Shcherbakov:2012appl} with the uncertainties being about a factor of $2$.
The correspondent ambient magnetic field strength is $B\approx0.05$~G in equipartition.
The equipartition argument is supported by a recent observational finding of a strong magnetic field at several arcseconds distance from Sgr A* \citep{Eatough:2013lz}.
Large magnetic field in the outer flow translates into the magnetic field energy density comparable to the thermal energy density in the inner flow, which was explicitly shown, e.g., in a spherical case
\citep{Scharlemann:1983wq,Shcherbakov:2008io}.

The corresponding perpendicular radius of the best-fitting magnetically arrested cloud is $\rho_p=1.9\times10^{14}{\rm cm}=150\rs$ according to the equation~(\ref{eq:rho}).
Then we estimate the normalization constant of the particle distribution to be a factor of $(\rho_{\rm Ram}/\rho_p)^2=28$ smaller.
The particle distribution with the slope $p=2.2$ does not continue till the infinite Lorentz factor, but breaks at a characteristic $\gamma_{\rm cool}$ determined by cooling \citep{Yuan:2003sg}.
The synchrotron cooling timescale is
\begin{equation}
t_{\rm synch}=\frac{6\pi m_e c}{\sigma_T \gamma B^2},
\end{equation} while the characteristic dynamical timescale is
\begin{equation}
t_{\rm dyn}\sim2t_{\rm ff}=4{\rm months},
\end{equation} so that the cooling break is at
\begin{equation}
\gamma_{\rm cool}\sim3\times10^4.
\end{equation}
The modified electron distribution is
\begin{equation}\label{eq:dist}
\frac{dN}{d\gamma}\approx 7\times10^{47}\cases{\gamma^{-2.2},& for $\gamma\le\gamma_{\rm cool}$\cr
\gamma_{\rm cool}\gamma^{-3.2}, & for $\gamma>\gamma_{\rm cool}$.}
\end{equation}
The peak frequency of the synchrotron emission is
\begin{equation}
\nu=\frac{3}{4\pi}\gamma^2 \frac{e B}{m_e c},
\end{equation} which corresponds to the frequency $\nu_{\rm cool}=2.7\times10^{14}$~Hz and the near-IR wavelength $\lambda_{\rm cool}=1.1\mu$m for the electrons at the cooling break.
Averaging the synchrotron emissivity \citep{Rybicki1979} over the pitch angles we obtain the specific luminosity
\begin{equation}\label{eq:Lnu}
L_\nu=7\times10^{47}\frac{\sqrt{3\pi}e^3 B}{2m_e c(p+1)}\left(\frac{2\pi m_e c\nu}{3e B}\right)^{-p/2+1/2}g,
\end{equation} where the constant equals $g=\Gamma(p/4+5/4)\Gamma(p/4+19/12)\Gamma(p/4-1/12)/\Gamma(p/4+7/4)$ with a gamma-function $\Gamma(x)$.
The optically thin specific flux is $F_\nu=L_\nu/(4\pi d^2)$, which gives
\begin{equation}
F_\nu\approx 0.38\left(\frac{\nu}{\rm GHz}\right)^{-0.6}~{\rm Jy}~{\rm for}~\lambda\ge1.1\mu{\rm m}
\end{equation} below the cooling break. This estimated flux is much lower than the prediction in \citet{Narayan:2012jl}.

The specific flux at $\nu=22$~GHz is expected to increase by only $\Delta F_\nu=0.06$~Jy, which is much lower than the quiescent level of Sgr A* at that frequency \citep{Shcherbakov:2012appl}.
The radio source would remain undetectable even at $\nu=1.4$~GHz \citep{Bower:2013za}.
The bow-shock $K$-band flux is
\begin{equation}
F_\nu(2\mu{\rm m})\approx 0.3{\rm mJy},
\end{equation} which is also much below the quiescent level of Sgr A* \citep{Dodds-Eden:2011jw}.
The maximum electron Lorentz factor
\begin{equation}
\gamma_{\rm max}=\left(\frac{6\pi e}{\sigma_T B}\right)^{1/2}\gtrsim10^8
\end{equation} is determined by the equality of the cooling time and the acceleration time \citep{Peer2005}.
Such high Lorentz factors are found in the supernova shock remnants, which appear to produce X-ray synchrotron emission \citep{Reynolds:1996qk}.
The electrons with relatively moderate Lorentz factors $\gamma\sim10^6$ produce the X-rays at the G2 pericenter.
The specific synchrotron flux above the cooling break behaves as $F_\nu\propto\nu^{-1.1}$ and the estimated intrinsic luminosity is
\begin{equation}
\nu L_\nu=2\times10^{33}\left(\frac{\nu}{4{\rm keV}}\right)^{-0.1}{\rm erg~s}^{-1}~{\rm for}~\lambda<1.1\mu{\rm m}.
\end{equation}
The predicted power $\nu L_\nu=2\times10^{33}$ at $\nu=4$~keV is slightly below the quiescent unabsorbed luminosity of Sgr A*
$L_{X}\sim3\times10^{33}{\rm erg~s}^{-1}$ \citep{Baganoff:2003,Wang:2013sc}, but is much below the luminosity $L_X=4\times10^{35}{\rm erg~s}^{-1}$ of the magnetar,
which turned on near Sgr A* on $2013$ April 23 \citep{Kennea:2013sg,Rea:2013ax}.

The bow shock should have passed through the pericenter around $2013$ March according to \citet{Sadowski:2013mn} ahead of the CM.
The specific flux increase by $\Delta F_\nu=0.4$~Jy at $22$~GHz was observed in $2013$ April \citep{Tsuboi:2013tel}, but
the observations in $2013$ June by \citet{Bower:2013za} showed that Sgr A* returned to the mean level of radio flux.
As the intrinsic variability of Sgr A* is about $0.4$~Jy at $22$~GHz \citep{Shcherbakov:2012appl}, then the rise observed in $2013$ April may have been unrelated to the cloud,
especially since no substantial flux increase was detected at the lower frequencies $\nu=1.5-14$~GHz \citep{Bower:2013za}.
No substantial X-ray flux increase was reported either from the Sgr A* region before the magnetar turned on in $2013$ April \citep{Neilsen:2013nh,Rea:2013ax}.

The ratio of the ambient magnetic field energy density to the radiation energy density is about $10^{2}$ at the pericenter.
Then the inverse Compton power is $100$ times lower than the synchrotron power for the same $\gamma$, and the particles cool via the synchrotron radiation.
The inverse Compton emission power \citep{Rybicki1979}
\begin{equation}
L_{C}=\frac43\sigma_T c\gamma^2  U_{\rm ph}
\end{equation} peaks at a frequency
\begin{equation}
\nu_C=\gamma^2\nu_{\rm ph},
\end{equation} where $\nu_{\rm ph}$ is a frequency of the seed optical and ultraviolet photons.
The inverse Compton X-rays are produced by the electrons with the Lorentz factors $\gamma=20-60$.
Convolution of the electron distribution with the photon field gives the inverse Compton X-ray luminosity
\begin{equation}
L_{X,C}(2-10{\rm keV})=4\times10^{29}{\rm erg~s}^{-1}
\end{equation}  much below the quiescent level of Sgr A*.
The inverse Compton scattering of the dust IR emission produces an even lower X-ray power, since the particle acceleration site is offset from the CM.

\subsection{Emission from the Bulk of the Cloud}
When the tidally disrupted object is not magnetized, then the tidal shock occurs at a specific location,
where the orbital planes of the independently moving cloud particles intersect \citep{Carter:1983aa,Luminet:1985yt}.
The shock can locally heat the gas up to the relatively high temperature, despite the compressed gas cools efficiently on the way to the pericenter \citep{Saitoh:2012co}.
However, the tidal shock does not occur in the magnetically arrested model, where the cloud is supported by the magnetic pressure.
Instead, gradual heating takes place over the entire volume of the cloud.

The best-fitting magnetically arrested model has the density $n_{\rm peri}=2.4\times10^6{\rm cm^{-3}}$,
the magnetic field $B_{\rm peri}\approx 0.7$~G, and the Alfven crossing time $t_{\rm A}\approx1$~mo at the pericenter.
The self-consistency of the model requires that the magnetic field dissipation time is much longer than the Alfven time $t_{\rm diss}\gtrsim 10t_{\rm A}=0.8$~yr.
Then the volume heating rate is
\begin{equation}
Q_{\rm vol}\lesssim\frac{B_{\rm peri}^2}{8\pi}t_{\rm diss}^{-1}\sim7\times10^{-10}{\rm erg~s}^{-1}{\rm cm^{-3}}.
\end{equation}
Heating is substantial only over a part of the cloud near the pericenter with the length $\Delta L\sim2R_p$, since $Q_{\rm vol}$ depends steeply on the distance to Sgr A*.

\citet{Gillessen:2012jq,Gillessen:2013pe} expect the interaction with the hot ambient medium to drive a strong shock into the cloud,
since the gas pressure of the cloud is much less than the ambient pressure. However, the ratio of the ambient gas pressure to the cloud magnetic pressure is
\begin{equation}\label{eq:pressures}
\frac{p_{\rm gas}}{p_{\rm magn}}=\frac{8\pi n_{\rm gas}k_B T_{\rm gas}}{B_{\rm peri}^2}\approx0.07
\end{equation} at the pericenter in the magnetically arrested model.
The ratio is much less than unity and the cloud is magnetically supported against the ambient hot gas.
Then we do not expect the shock or the corresponding X-ray radiation.
The equation~(\ref{eq:pressures}) also justifies neglecting the ambient gas force $F_{\rm out}$ in the force balance given by the equation~(\ref{eq:fbalance}).

We simulate the emission from the part of the cloud near the pericenter heated by the magnetic field dissipation.
We take the perpendicular size and cloud density based on formulas (\ref{eq:rho}) and (\ref{eq:magdense}), respectively, at a distance $r=r_{\rm peri}$.
Extra heating with a specified rate per unit volume is a standard option in CLOUDY, while the gas temperature is a standard code output.
The model with the dissipation time $t_{\rm diss}=10t_{\rm A}$ produces relatively warm gas at $T\approx5\times10^4$~K, which emits less powerful hydrogen and helium lines.
The dust temperature reaches $T_{\rm dust}\approx800$~K, which is still below the sublimation threshold \citep{Guhathakurta:1989al,Ferland:2013cp}.
Such dust manifests as an IR source with the apparent magnitudes $m_M=13.4$, $m_{L'}=14.0$, and $m_{K_s}=17.2$, which are comparable in $M$ and $L'$ band to $2011$ data and
up to $3$~mag brighter than the upper limit in $K_s$ band.
The gas cools more efficiently for the larger dissipation times $t_{\rm diss}\gtrsim20t_{\rm A}$, so that the gas temperature settles to $8,000<T<4\times10^4$~K.
This drives Br$\gamma$ luminosity down from the case of no extra heating.
If only the ionizing radiation heats the gas, then the simulated Br$\gamma$ luminosity is $L({\rm Br}\gamma)\sim2\times10^{31}{\rm erg~s}^{-1}$ at the pericenter
due to the larger cloud density and the higher S0-2 flux, which is $2$ times brighter than the observed level.
Even small heating with $t_{\rm diss}\gtrsim20t_{\rm A}$ decreases the Br$\gamma$ emissivity in the optically thin models,
which could lead to a constant or a decreasing Br$\gamma$ flux from the cloud as it passes through the pericenter.
Non-equilibrium ionization effects may as well modify pericenter Br$\gamma$ luminosity.

\section{DISCUSSION AND CONCLUSIONS}\label{sec:discussion}
In the present paper we report the line and the dust diagnostics of the G2 object, hypothesized to be a gas cloud, moving towards Sgr A* in the Galactic Center.
We consider three cloud shapes: the spherical, tidally distorted, and magnetically arrested.
The cloud shape might be close to spherical in the presence of the central star.
The tidally distorted shape is appropriate for the unmagnetized cloud without the central object.
The magnetically arrested regime is representative of the magnetized cloud behavior.
We identify the optically thin and the optically thick groups of models based on the optical depth to the incident ionizing radiation.
The models perform differently, when compared to the data. The optically thin spherical models show the constant Br$\gamma$ luminosity,
despite \citet{Scoville:2013ql} expects only the collisionally ionized gas with the internal energy source to reproduce $L({\rm Br}\gamma)=\rm const$.
However, the spherical models overestimate the cloud size above the observed value.
The optically thin tidally distorted shapes show a relatively large increase of $L({\rm Br}\gamma)$ inconsistent with the observations,
while the optically thick tidally distorted shapes provide worse spectral fits.
All computed magnetically arrested models show the relatively weak increase of $L({\rm Br}\gamma)$ with time marginally consistent with the observations.

The best-fitting magnetically arrested model has the initial density $n_{\rm init}=1.8\times10^5{\rm cm}^{-3}$, the initial radius $R_{\rm init}=2.2\times10^{15}{\rm cm}=17 {\rm mas}$,
the cloud mass $m_{\rm cloud}=4.1M_{\rm Earth}$, and the dust abundance $A=0.072$ relative to the ISM. Such cloud forms as a spherical object at the distance $r_{\rm init}=1$~arcsec from Sgr A*.
From the place of formation to the pericenter the cloud is in the perpendicular balance of the magnetic force and the gravitational force.
This model reaches a good agreement with the Br$\gamma$ luminosity, the ratio $L({\rm He}I)/L({\rm Br}\gamma)$, and
the $L'$ and $M$ magnitudes observed in $2011$. It is marginally consistent with the luminosities and the magnitudes reported in $2004$ and $2008$.
The radial stretching of the cloud leads to the projected size in agreement with the observed spatial extent in $2008$ and $2011$, while
the correspondent spread of the radial velocity is underpredicted. Simultaneous agreement of the velocity spread and the cloud size may require resolution
of the observational uncertainties or the modeling approximations.

The $L'$ band emission is simulated to grow by $\Delta M_{L'}=0.5$~mag from $2004$ till $2011$ and is expected to reach $m_{L'}=13.5$~mag at the pericenter as the incident flux further increases
by $40\%$ compared to $2011$ (see Table~\ref{tab:obs}). This is comparable to the brightness produced due to the magnetic field dissipation.
The dim S stars provide an additional contribution to the incident flux at the pericenter
due to the rising inwards surface brightness profile $\Sigma\propto r^{-0.93\pm0.09}$ \citep{Do:2013kl}.
They are not expected to produce a substantial number of the ionizing photons, but their radiation is reprocessed by the dust.
However, the internal dust heating may dominate the heating by the incident flux.
The dissipation of the magnetic energy near the pericenter raises the temperature of the gas, which collisionally heats the dust.
As a result, the dust emits substantially more IR light and the cloud could reach total observed magnitude up to $m_{L'}=13$.
The observations of the pericenter passage in the IR band probe the magnetic energy dissipation rate,
though such observations are difficult due to the source confusion near Sgr A* \citep{Phifer:2013ap}.

The bow shock region may produce a distinct radiative signature.
The bow shock should brighten in the radio band and the X-rays some time before the center of mass passes through the pericenter.
The brightening lasts for the dynamical time $t_{\rm dyn}\sim4$~mo at the wavelengths $\lambda<1.1\mu$m affected by the synchrotron cooling.
However, the best-fitting magnetically arrested model predicts the radio and X-ray fluxes below the current quiescent level of Sgr A*.
The radio flux is a factor of $30$ lower than the prediction in \citet{Narayan:2012jl} owing mainly to the smaller cross-section.
Yet, the predicted emission is still uncertain by a factor of several, since the particle acceleration efficiency is not known \citep{Narayan:2012jl}.
The positive detection or the non-detection of the radio flux from G2 constrain the product of the particle acceleration efficiency by the bow shock cross-section.
The X-ray observations constrain the acceleration of particles to the large Lorentz factors $\gamma\sim10^6$.
It is instructive to compare the presented dusty cloud emission modeling with the scenario, where the recombination lines are emitted within the wind from a massive central star \citep{Scoville:2013ql}.
The latter scenario naturally produces very low radio flux from the bow shock owing to a very small bow shock cross-section \citep{Crumley:2013vl}.
However, Br$\gamma$ luminosity may grow substantially with time in this scenario \citep{Ballone:2013bv} in contradiction with the observations.

Following the pericenter passage the cloud is disrupted by the combined effects of the Kelvin-Helmholtz and Rayleigh-Taylor instabilities,
the external pressure, the differential gravitational force, and the conductive evaporation of the cloud \citep{Gillessen:2012jq,Burkert:2012ca,Anninos:2012kz,Saitoh:2012co,Ballone:2013bv}.
We show that the cloud mass may be much larger $m_{\rm cloud}=(4-20)M_{\rm Earth}$ compared to the estimate in \citet{Gillessen:2012jq}.
The resultant BH accretion rate and the accretion flow luminosity may also be higher \citep{Moscibrodzka:2012ax}, if the disruption dynamics is unchanged.
The heavier clouds are to be tested with the future numerical simulations of G2.

Despite the agreement with the data, the best-fitting magnetically arrested model might not fully represent the G2 cloud.
We considered three idealized cases for the shape, while more options are possible. If the magnetization is relatively low $\sigma<1$, then
the cloud starts off as the tidally distorted shape, while closer to the pericenter it switches to the magnetically arrested regime.
However, if the switching happens after $2004$, then such hybrid models may exhibit a large rise of the Br$\gamma$ luminosity inconsistent with the observations.
The magnetized cloud might not closely follow the computed magnetically arrested shape.
If the magnetic field in the cloud dissipates on the dynamical timescale, then the object approaches the denser tidally distorted shape.
Such models are similarly disfavored by the observations. Finally, the dynamical effects acting on the cloud may reduce its mass.
Then $L({\rm Br}\gamma)$ could rise less steeply or even be constant with time in the magnetically arrested models, so that better consistency with the observations is reached.

\section{ACKNOWLEDGEMENTS}\label{sec:acknowledgements}
This work is supported by NASA Hubble Fellowship grant HST-HF-51298.01 (RVS).
The author is thankful to Frederick Baganoff, Kazimierz Borkowski, Stephen Reynolds, and Chris Reynolds for stimulating discussions, with
special thanks to Stefan Gillessen and Tobias Fritz for discussions of the observations and extensive feedback.
The author acknowledges hospitality of the Physics and Astronomy Department, University of North Carolina, Chapel Hill, where a part of the work was conducted.
\bibliographystyle{apj}

\end{document}